\documentclass[12pt]{article}

\newcommand{\be}{\begin{equation}}
\newcommand{\ee}{\end{equation}}
\newcommand{\bea}{\begin{eqnarray}}
\newcommand{\eea}{\end{eqnarray}}
\newcommand{\p}{\partial}
\newcommand{\fr}{\frac}
\newcommand{\I}{\mathcal{I}}

\begin{document}


\begin{titlepage}

\title{On the Hilbert space dimension of systems with second class constraints}

\author{M.N. Stoilov\\
{\small\it Bulgarian Academy of Sciences,}\\
{\small\it Institute of Nuclear Research and Nuclear Energy,}\\
{\small\it Blvd. Tzarigradsko Chausse\'e 72, Sofia 1784, Bulgaria}\\
{\small e-mail: mstoilov@inrne.bas.bg}}

\maketitle

\begin{abstract}
It is shown that quantized dynamical system with second class constraints has infinite dimensional Hilbert space.
\end{abstract}

{\it
\vskip 10pt

}

\end{titlepage}


The quantization of a classical dynamical system, considered as a mathematical problem, is a map
from  the real functions on the system phase space to
self-adjoin operators in some Hilbert space.
The Hilbert space can be with finite or infinite dimension 
depending on the  model we quantize.
For example, if we have a flat phase space with globally separated coordinates and momenta 
then the resulting Hilbert space is infinite dimensional. 
On the other hand, if we have a compact  phase space, 
then the Hilbert space is finite dimensional \cite{H}. 
Here we shall consider the quantization of systems with second class constraints, 
mainly the question about the dimension of the corresponding Hilbert space.
Any system of this type exhibits properties which allow us to think that it  can interpolate
between models with flat and compact phase spaces:
 it is defined on a flat phase space but this space is larger than the real one; 
the real phase space can be very complicated and 
with highly non trivial analog of the Poisson bracket on it 
thus resembling a system with compact phase space.
Therefore, it will be instructive to determine the dimension 
of the Hilbert space of a quantized system with second class constraints.

We start with a simple demonstration how one can  deduce whether the dimension of the Hilbert space is finite or infinite depending on the quantized system.

Consider a model with flat phase space $R^{2n}$. Let $x_i, i=1, \dots , n$ are the coordinates and
$p_i, i=1,\dots, n$ are  corresponding momenta.
Let $\omega=\sum_i dx_i \land dp_i$ is the canonical symplectic form on $R^{2n}$.
Using this form we define the Poisson bracket between $(C^\infty)$ functions on the phase space  
\be
\{f(x,p),g(x,p)\}_{P.B.}=\fr{\p f}{\p x_i} \fr{\p g}{\p p_i} -
\fr{\p f}{\p p_i} \fr{\p g}{\p x_i}. \label{one} 
\ee
In particular from Eq.(\ref{one}) we have 
\be
\{ x_i, p_j\}_{P.B.} =  \delta_{ij}. \label{two}
\ee
In the classical case eqs.(\ref{one}) and (\ref{two}) are equivalent.
It is not the same when we quantize the system.
 
During the quantization  we cannot map consistently all functions on the phase space 
to self-adjoint operators in a Hilbert space.
But we can a map a  set of functions  $\{f\}$,  
which is as large as possible and forms a closed algebra under Poisson brackets,
to a set  $\{\hat f\}$ of self-adjoint operators. 
The functions  $\{ f\}$ are called  primary quantities 
and the correspondence $ f \rightarrow \hat f$ has  the following properties:
\begin{enumerate} 
\item The constants are primary quantities and  $1\rightarrow \I$ 
where $\I$ is the identity operator in the Hilbert space.
\item The operator image of the Poisson bracket 
is the commutator of the corresponding operators: 
\be
\widehat{ \{f, g\}}_{P.B.} = i \hbar [\hat f, \hat g ],
\ee
i.e.  $\{\hat f\}$ is a Lie algebra representation of $\{f\}$.
\end{enumerate}
In the flat phase space example which we  are considering now  the primary quantities 
can be either linear functions of momenta and arbitrary functions of coordinates \cite{K}, 
or quadratic polynomials of coordinates and momenta \cite{GW}. 
In both cases we have
\be
[\hat x_i,\hat  p_j ] = i\hbar  \delta_{ij} \I \label{star}
\ee
(plus other commutation relations depending on what is our choice for $\{f\}$.)
So, according to eq.({\ref{star}), we have a representation  of the  Heisenberg algebra in the Hilbert space.
This fact allows us to demonstrate that  the Hilbert space is with infinite dimension.
Indeed, if it is with finite dimension $D$ then, 
taking the trace  of both sides of eq.(\ref{star}) we will obtain a contradiction
$0 = i \hbar \delta_{ij} D$.

Consider now a dynamical system which is symmetric with respect to the action of some Lie group. 
The symmetry acts by definition as canonical transformations 
and its algebra has a representation in the functions on the phase space 
which is closed under time evolution, i.e.
\bea
\{g_a, g_b\}_{P.B.} & = & c_{abc} g_c \label{alg}\\
\{g_a, H\}_{P.B.} & = & h_{ab}g_b,  \label{alh}
\eea
where $g_a$ are the generators of the Lie algebra and $H$ is the Hamiltonian of the system.
If we want to quantize such system and we are only interested in the symmetry, 
e.g.  spin quantization, Chern-Simon theories, and in general, 
quantization on co-adjoint orbits of Lie groups, 
then it is natural to use symmetry generators as primary quantities
and  eventually the Hamiltonian 
if it is not zero or is not a combination of the symmetry generators.
In this case the quantization of eqs.(\ref{alg}, \ref{alh}) will look as follows
\bea
\left[\hat g_a,\hat g_b \right] & = & i \hbar c_{abc} \hat g_c \label{qalg}\\
\left[\hat g_a,\hat H \right] & = & i \hbar h_{ab} \hat g_b. \label{qalh}
\eea
If the  symmetry algebra can be represented with trace-less matrices,
e.g. it is a simple Lie algebra, then 
the same arguments which show that the representations of the Heisenberg algebra are infinite   
dimensional lead us to the conclusion that a system which primary quantities satisfy 
eqs.(\ref{qalg}, \ref{qalh}) can have a finite dimensional Hilbert space. 
In this case the primary quantities are mapped into constant matrices.
This is exactly the reason why we can use Pauli matrices as the electron spin operators.

The above examples demonstrate that the dimension of the Hilbert space of a quantized system depends 
on the algebra which we realize in it. 
Now we turn our attention to the main subject of this note, 
namely to  systems with second class constraints.
As we have mentioned already these systems can have very complicated 
analog of Poisson brackets even between canonical coordinate and momenta.

Let the phase space of our model is again $R^{2n}$ but now it is subjected to $2m$  second class constraints $\chi_a, a=1,\dots, 2m$.
Second class constraints are always even number and 
always $m \le n$.
The case $m = n$ corresponds to a trivial system with no dynamical degrees of freedom.

Two simple examples explain why we can think about systems with second class constraints as
interpolating between systems with flat and compact phase space.
First, let   $m=1$ and $\chi_1=x_1;\;\chi_2=p_1$.
In this case the system with second class constraints is equivalent to 
a standard system in $R^{2(n-1)}$
where the first coordinate and its momentum are omitted.
Therefore the dimension of the Hilbert space of this model is infinite.
Second, let again   $m=1$ and
but now $\chi_1=x_i x^i-r^2;\;\chi_2=p_i x^i$.
This system describes a model with compact configuration space.
The question we want to address  is whether it is possible to find 
a system which after quantization to have a finite dimensional Hilbert space.

When we work with a system with second class constraints
then the usual Poisson bracket is replaced by Dirac one \cite{D}:
\be
\{f, g\}_D = \{f, g\}_{P.B.} - \{f, \chi_a\}_{P.B.}
 \Delta_{ab}^{-1} \{\chi_b, g\}_{P.B.}. \label{three}
\ee
Here $\Delta_{ab}=\{\chi_a, \chi_b\}_{P.B.}$.
The matrix $\Delta$ is always invertible 
(one can think that this is a definition of system with second class constraints),
so it is correct to use $\Delta^{-1}$ in eq.(\ref{three}).

The quantization of systems with second class constarints follows the same rules 
as standard quantization, but now everywhere the Poisson brackets are replaced
by Dirac ones.
Choosing $x$ and $p$ as primary quantities we want to understand if 
there is a Heisenberg sub-algebra is their commutation relations.
For this purpose we consider the quantity $\{x_m, p^m\}_D$
\be
\{x_i, p^i\}_D = n - \{\chi_b,p^i\}_{P.B.}\{x_m, \chi_a\}_{P.B.} \Delta_{ab}^{-1} =
n-m
\ee
where we have used 
$\{\chi_b,p^i\}_{P.B.}\{x_i, \chi_a\}_{P.B.}=\p \chi_b/\p x_i  \p \chi_a/\p p^i$
 and the skew-symmetry of the matrix $\Delta$.
Therefore,  in any case  there is  a Dirac bracket 
which is a number like in the case of usual Poisson brackets.
In this way we have proved the following  lemma:

{\bf Lemma:} 
The Hilbert space of any Bose-Einstein quantized dynamical system 
with second class constraints is infinite dimensional.


\begin{thebibliography}{9.}
\bibitem{H} N. Hitchin, Comm.Math.Phys. 131(2) (1990) 347-380.
\bibitem{K}A.A.  Kirilov 
{\it Elements of the Theory of Representations, Springer-Verlag}, 
Berlin Heidelberg New York 1976
\bibitem{GW} S. Gukov, E. Witten, ATMP {\bf 13} (2009) 1-73.
\bibitem{D} P.A.M. Dirac,
{\it Lectures on Quantum Mechanics},
Yeshiva Univ. (Academic Press, New York, 1967).
%
\end{thebibliography}
\end{document}